\documentclass[runningheads]{llncs}
\usepackage{graphicx}
\usepackage{array} 
\usepackage{arydshln} 
\usepackage[numbers,sort&compress]{natbib}

\usepackage{tabularray}

\usepackage{caption}
\usepackage{subcaption}

\usepackage[utf8]{inputenc}
\usepackage[commandnameprefix=ifneeded,commentmarkup=uwave]{changes}
\usepackage[n,landau,notions,ff,mm]{cryptocode}
\usepackage{xcolor}
\usepackage{graphicx}
\definecolor{gamechangecolor}{gray}{0.74}
\usepackage{tikz}
\usetikzlibrary{arrows,positioning,decorations.pathreplacing}
\usepackage{url}
\usetikzlibrary{shapes}
\usepackage{import}
\usepackage{xifthen}
\usepackage{pdfpages}
\usepackage{transparent}
\usepackage{balance}
\usepackage{xspace}
\usepackage{amsmath}
\usepackage{hyperref}
\usepackage{xurl}
\usepackage{cleveref}
\usepackage{amsfonts}%
\usepackage{multirow}
\usepackage{graphicx}
\usepackage{booktabs}
\usepackage{caption}
\usepackage{enumitem}
\setlist[itemize]{left=0pt}
\setlist[enumerate]{left=0pt}
\usepackage{makecell}

\usepackage{todonotes}
\setuptodonotes{inline, color=blue!30}
\usepackage[framemethod=TikZ]{mdframed}
\mdfsetup{
    font=\small,
    leftmargin=5pt,
    rightmargin=5pt,
    skipbelow=5pt,
    skipabove=5pt,
    innertopmargin=5pt,
    innerbottommargin=5pt,
    innerleftmargin=5pt,
    innerrightmargin=5pt,
    frametitlerule=true,
    frametitlealignment=\hspace*{0pt}
}

\usepackage{tikz-qtree}
\usepackage{tikz}
\usepackage{pgf-umlsd}
\usepackage{dashbox}

\usepgflibrary{arrows} 
\usepackage{pgfplots}


\newcommand{\etal}{\textit{et al.}\xspace}

\definecolor{cblue}{rgb}{0.0, 0.28, 0.9}
\definechangesauthor[name=duc, color=cblue]{duc}
\definecolor{zp}{rgb}{0.0, 0.28, 0.4}
\definechangesauthor[name=zhipeng, color=zp]{zhipeng}

\definecolor{forest}{rgb}{0.0, 0.5, 0.0}

\newcommand{\UST}{$\texttt{UST}$\xspace}
\newcommand{\LUNA}{$\texttt{LUNA}$\xspace}

\definecolor{Gray}{gray}{0.9}

\usepackage{amssymb}
\usepackage{pifont}

\usepackage{acro}
\acsetup{single}

\DeclareAcronym{SEC}{
  short = SEC,
  long  = Securities and Exchange Commission,
}
\newcommand{\SEC}{\ac{SEC}\xspace}

\DeclareAcronym{CFTC}{
  short = CFTC,
  long  = Commodity Futures Trading Commission,
}
\newcommand{\CFTC}{\ac{CFTC}\xspace}

\DeclareAcronym{FCA}{
  short = FCA,
  long  = Financial Conduct Authority,
}
\newcommand{\FCA}{\ac{FCA}\xspace}

\DeclareAcronym{PSRs}{
  short = PSRs,
  long  = Payment Service Regulations,
}
\newcommand{\PSRs}{\ac{PSRs}\xspace}

\DeclareAcronym{EMRs}{
  short = EMRs,
  long  = Electronic Money Regulations,
}
\newcommand{\EMRs}{\ac{EMRs}\xspace}

\DeclareAcronym{AML}{
  short = AML,
  long  = Anti-Money Laundering,
}
\newcommand{\AML}{\ac{AML}\xspace}

\DeclareAcronym{CTF}{
  short = CTF,
  long  = Counter-Terrorist Financing ,
}
\newcommand{\CTF}{\ac{CTF}\xspace}

\DeclareAcronym{BoE}{
  short = BoE,
  long  = Bank of England,
}
\newcommand{\BoE}{\ac{BoE}\xspace}

\DeclareAcronym{FSMA}{
  short = FSMA,
  long  = Financial Services and Markets Act,
}
\newcommand{\FSMA}{\ac{FSMA}\xspace}

\DeclareAcronym{MiFiD}{
  short = MiFiD II,
  long  = Markets in Financial Instruments Directive II,
}
\newcommand{\MiFiD}{\ac{MiFiD}\xspace}

\DeclareAcronym{AMLD}{
  short = AMLD V,
  long  = Anti-Money Laundering Directive V,
}
\newcommand{\AMLD}{\ac{AMLD}\xspace}

\DeclareAcronym{PSD}{
  short = PSD II,
  long  = Payment Services Directive II,
}
\newcommand{\PSD}{\ac{PSD}\xspace}

\DeclareAcronym{CASP}{
  short = CASP,
  long  = Cryptoasset Service Provider,
}

\newcommand{\CASPs}{\acp{CASP}\xspace}

\DeclareAcronym{RAO}{
  short = RAO,
  long  = Regulated Activities Order,
}
\newcommand{\RAO}{\ac{RAO}\xspace}

\DeclareAcronym{FPO}{
  short = FPO,
  long  = Financial Promotion Order,
}
\newcommand{\FPO}{\ac{FPO}\xspace}

\DeclareAcronym{TFR}{
  short = TFR,
  long  = Transfers of Funds Regulation,
}
\newcommand{\TFR}{\ac{TFR}\xspace}

\DeclareAcronym{SEA}{
  short = SEA,
  long  = Securities Exchange Act,
}
\newcommand{\SEA}{\ac{SEA}\xspace}

\DeclareAcronym{CEA}{
  short = CEA,
  long  = Commodity Exchange Act,
}
\newcommand{\CEA}{\ac{CEA}\xspace}

\DeclareAcronym{PBC}{
  short = PBC,
  long  = People's Bank of China,
}
\newcommand{\PBC}{\ac{PBC}\xspace}

\DeclareAcronym{RBI}{
  short = RBI,
  long  = Reserve Bank of India,
}
\newcommand{\RBI}{\ac{RBI}\xspace}

\DeclareAcronym{NCA}{
  short = NCA,
  long  =  National Competent Authority,
}
\newcommand{\NCA}{\ac{NCA}\xspace}

\DeclareAcronym{HMT}{
  short = HMT,
  long  = HM Treasury,
}
\newcommand{\HMT}{\ac{HMT}\xspace}

\DeclareAcronym{OFAC}{
  short = OFAC,
  long  = Treasury's Office of Foreign Assets Control,
}
\newcommand{\OFAC}{\ac{OFAC}\xspace}

\DeclareAcronym{PSA}{
  short = PSA,
  long  = Payment Services Act,
}
\newcommand{\PSA}{\ac{PSA}\xspace}

\DeclareAcronym{MiCA}{
  short = MiCA,
  long  = Markets in Crypto-Assets,
}
\newcommand{\MiCA}{\ac{MiCA}\xspace}

\DeclareAcronym{EMD}{
  short = EMD,
  long  = Electronic Money Directive,
}
\newcommand{\EMD}{\ac{EMD}\xspace}

\DeclareAcronym{CATF}{
  short = CATF,
  long  = Cryptoassets Taskforce,
}

\DeclareAcronym{ICO}{
  short = ICO,
  long  = Initial Coin Offering,
}

\newcommand{\ICOs}{\acp{ICO}\xspace}

\DeclareAcronym{TC}{
  short = TC,
  long  = Tornado Cash,
}
\newcommand{\TC}{\ac{TC}\xspace}

\DeclareAcronym{DeFi}{
  short = DeFi,
  long  = Decentralized Finance,
}
\newcommand{\DeFi}{\ac{DeFi}\xspace}

\DeclareAcronym{DLT}{
  short = DLT,
  long  = Distributed Ledger Technology,
}
\newcommand{\DLT}{\ac{DLT}\xspace}

\DeclareAcronym{PoW}{
  short = PoW,
  long  = Proof-of-Work,
}

\DeclareAcronym{PoS}{
  short = PoS,
  long  = Proof-of-Stake,
}

\DeclareAcronym{LTVV}{
  short = LTV,
  long  = Loan-to-Value,
}

\DeclareAcronym{HF}{
  short = HF,
  long  = Health Factor,
}

\DeclareAcronym{LT}{
  short = LT,
  long  = Liquidation Threshold,
}

\DeclareAcronym{APR}{
  short = APR,
  long  = Annual Percentage Rate,
}

\DeclareAcronym{ROI}{
  short = ROI,
  long  = Return On Investment,
}

\DeclareAcronym{JIT}{
  short = JIT,
  long  = Just-in-Time,
}

\DeclareAcronym{NO}{
  short = NO,
  long  = Node Operator,
}

\DeclareAcronym{LSD}{
  short = LSD,
  long  = Liquid Staking Derivative,
}

\DeclareAcronym{LST}{
  short = LST,
  long  = Liquid Staking Tokens,
}

\DeclareAcronym{DEX}{
  short = DEX,
  long  = Decentralized Exchange,
}

\newcommand{\DEXes}{\acp{DEX}\xspace}

\DeclareAcronym{CEX}{
  short = CEX,
  long  = Centralized Exchange,
}

\newcommand{\CEXes}{\acp{CEX}\xspace}

\DeclareAcronym{EOA}{
  short = EOA,
  long  =  Externally-Owned Account,
}

\DeclareAcronym{NFT}{
  short = NFT,
  long  = Non-fungible Token,
}
\newcommand{\NFT}{\ac{NFT}\xspace}
\newcommand{\NFTs}{\acp{NFT}\xspace}

\DeclareAcronym{LP}{
  short = LP,
  long  = Liquidity Provider,
}

\DeclareAcronym{PtP}{
  short = PtP,
  long  = peer-to-peer,
}
\newcommand{\PtP}{\ac{PtP}\xspace}

\DeclareAcronym{TVL}{
  short = TVL,
  long  = Total Value Locked,
}
\newcommand{\TVL}{\ac{TVL}\xspace}

\DeclareAcronym{CFMM}{
  short = CFMM,
  long  = Constant Function Market Maker,
}

\DeclareAcronym{CPMM}{
  short = CPMM,
  long  = Constant Product Market Maker,
}

\DeclareAcronym{DApp}{
  short = DApp,
  long  = Decentralized Application,
}

\DeclareAcronym{DAO}{
  short = DAO,
  long  = Decentralized Autonomous Organization,
}
\newcommand{\DAO}{\ac{DAO}\xspace}

\DeclareAcronym{CeFi}{
  short = CeFi,
  long  = Centralized Finance,
}

\DeclareAcronym{MEV}{
  short = MEV,
  long  = Miner Extractable Value,
}

\DeclareAcronym{EV}{
  short = EV,
  long  = Extractable Value,
}

\DeclareAcronym{BEV}{
  short = BEV,
  long  = Blockchain Extractable Value,
}

\DeclareAcronym{AMM}{
  short = AMM,
  long  = Automated Market Maker,
}

\DeclareAcronym{FaaS}{
  short = FaaS,
  long  = Front-running as a Service,
}

\DeclareAcronym{SaaS}{
  short = SaaS,
  long  = Staking as a Service,
}

\DeclareAcronym{HFT}{
  short = HFT,
  long  = High-frequency Trading,
}

\DeclareAcronym{PGA}{
  short = PGA,
  long  = Priority Gas Auction,
}

\DeclareAcronym{BRF}{
  short = BRF,
  long  = Back-run Flooding,
}

\DeclareAcronym{PRG}{
  short = PRG,
  long  = Priority Gas Auction,
}

\DeclareAcronym{AE}{
  short = AE,
  long  = Atomic Execution,
}

\DeclareAcronym{BEET}{
  short = BEET,
  long  = Break-even Extraction Threshold,
}

\DeclareAcronym{BAD}{
  short = BAD,
  long  = Breaking Atomicity and Determinism,
}

\DeclareAcronym{aamm}{
  short = A$^2$MM,
  long  = Automated Arbitrage Market Maker,
}

\DeclareAcronym{dfmm}{
  short = DFMM,
  long  = Dynamic Fee Market Maker,
}

\DeclareAcronym{dfaamm}{
  short = A$^2_F$MM,
  long  = Automated Arbitrage and Fee Market Maker,
}

\DeclareAcronym{MVI}{
  short = MVI,
  long  = Minimum Victim Input,
}

\DeclareAcronym{BSC}{
  short = BSC,
  long  = Binance Smart Chain,
}

\DeclareAcronym{KS}{
  short = KS,
  long  = Kolmogorov-Smirnov,
}

\newcommand{\FTT}{\ensuremath{\xspace\texttt{FTT}}\xspace}

\newcommand{\BTC}{\ensuremath{\xspace\texttt{BTC}}\xspace}
\newcommand{\USD}{\ensuremath{\xspace\texttt{USD}}\xspace}
\newcommand{\USDT}{\ensuremath{\xspace\texttt{USDT}}\xspace}

\newcommand{\XRP}{\ensuremath{\xspace\texttt{XRP}}\xspace}

\newcommand{\ETH}{\texttt{ETH}\xspace}

\begin{document}
\title{Global Trends in Cryptocurrency Regulation: An Overview}

%

\author{Xihan Xiong\inst{1}\and Junliang Luo\inst{2}}

\authorrunning{Xiong et al.}
\institute{Imperial College London \and McGill University}

\maketitle             

\begin{abstract}
Cryptocurrencies have evolved into an important asset class, providing a variety of benefits. However, they also present significant risks, such as market volatility and the potential for misuse in illegal activities. These risks underline the urgent need for a comprehensive regulatory framework to ensure consumer protection, market integrity, and financial stability. Yet, the global landscape of cryptocurrency regulation remains complex, marked by substantial variations in regulatory frameworks among different countries. This paper aims to study these differences by investigating the regulatory landscapes across various jurisdictions. We first discuss regulatory challenges and considerations, and then conduct a comparative analysis of international regulatory stances, approaches, and measures. We hope our study offers practical insights to enhance the understanding of global trends in cryptocurrency regulation.
\end{abstract}

\section{Introduction}
Cryptocurrencies, since their inception with Bitcoin in $2009$, have emerged as a revolutionary asset class providing an alternative to traditional fiat currencies. Their advantages include enhanced transactional transparency, reduced processing times, and increased accessibility to financial services, particularly in underbanked regions. Over the years, the cryptocurrency market has experienced exponential growth, evolving into a significant component of the global financial ecosystem with a market capitalization of $1.6$t \USD\footnote{\url{https://coinmarketcap.com/charts/}, accessed on Jan~$20$,~$2024$.}. This expansion reflects the growing integration of cryptocurrencies into the broader economy. However, despite their increasing popularity and the innovation they bring to financial services and beyond, cryptocurrencies are not without their risk. Major economies such as the European Union (EU)\footnote{See \href{https://www.esma.europa.eu/press-news/esma-news/eu-financial-regulators-warn-consumers-risks-crypto-assets}{EU financial regulators warn consumers on the risks of crypto-assets}.} have issued warnings about the potential risks. 
These concerns range from their use in illicit activities, such as money laundering and financing of terrorism, to the risks of tax evasion, market manipulation, and market volatility. Such risks underscore the complex landscape within which cryptocurrencies operate, balancing their potential for innovation against the need for consumer protection and financial stability.

Given these considerations, understanding how different countries implement the regulatory framework for cryptocurrencies becomes crucial. Currently, the global regulatory landscape remains fragmented, with countries adopting stances ranging from outright bans to recognizing cryptocurrencies as legal tender.

In this paper, we explore the global regulatory landscape for cryptocurrencies, assessing the approaches implemented by countries worldwide. The main contributions of this paper are outlined as follows: 

\vspace{+1mm}
\noindent $\bullet$ We identify the principal challenges in 
regulating cryptocurrencies that arise from their characteristics and risks inherent to the underlying blockchain technology. We also discuss key considerations that necessitate regulatory attention.  
 
\vspace{+0.5mm}
\noindent$\bullet$~ We examine the cryptocurrency regulatory frameworks across all countries and propose a comprehensive taxonomy to classify the regulatory stances, approaches, and measures implemented by countries worldwide. Through our analysis, we observe significant variations in regulatory progress and find that cryptocurrencies remain unregulated in at least $71$ countries at the time of writing.

\vspace{+0.5mm}
\noindent$\bullet$~ We explore how various jurisdictions are navigating the challenges and opportunities presented by cryptocurrencies. Differences in regulatory approaches among key global players are highlighted, along with their implications. 



\section{Related Work}\label{sec:related_work}
The regulation of cryptocurrencies has received extensive attention within the academic community. Numerous studies offer an overview of cryptocurrency regulation. For example, Blandin~\etal~\cite{blandin2019global} provided a comparative analysis of the current regulatory landscape of cryptoassets in 23 jurisdictions. Cumming~\etal~\cite{cumming2019regulation} analyzed the \SEC's initial and subsequent statements on \ICOs to highlight the challenges of applying existing legal frameworks to a rapidly evolving crypto space. Meizquita~\etal~\cite{mezquita2023cryptocurrencies} studied the diverse perspectives and interests that influence each country's decision-making process in adopting specific regulatory frameworks. The primary distinction between this study and earlier ones lies in the scope of analysis. While the earlier studies selectively focused on regulations in specific countries, our research extends to examining cryptocurrency regulatory frameworks across all countries. Additionally, we develop a comprehensive taxonomy to systematically categorize and compare the regulatory stances, approaches, and measures implemented by countries worldwide.
%

\section{Background}\label{sec:background}

    \subsection{Blockchain, DLT and DeFi}

    Blockchain is a decentralized ledger technology that records transactions across multiple nodes to ensure the integrity and security of a data record without the need for a central authority. It organizes data into consecutive blocks that are cryptographically linked and immutable, thereby providing a verifiable transaction history. As of the latest update, Ethereum stands as the leading blockchain platform, boasting a \TVL of $30$b \USD\footnote{\url{https://defillama.com/chain/Ethereum}, last accessed on Jan~$11$,~$2024$.}.

     \DLT represents a broader category. Indeed, a universally accepted definition of \DLT does not exist\footnote{See section 2.9 of \href{https://www.fca.org.uk/publication/consultation/cp19-03.pdf}{CP19/3: Guidance on Cryptoassets}.}. Broadly, \DLT refers to a digital system that records asset transactions across multiple locations simultaneously without a centralized authority or data storage. \DLT enables \PtP transactions and supports diverse applications beyond cryptocurrencies, such as supply chain management and identity verification.

    \DeFi~\cite{werner2022sok} is a financial system built on blockchain technology that allows for financial transactions without centralized intermediaries, using smart contracts on networks such as Ethereum. Cryptocurrencies are integral to \DeFi, serving both as assets and as mechanisms for various activities, including lending, borrowing, and \DEXes trading.
    
    \subsection{Cryptocurrency Primer}

    \noindent\textbf{Definitions}
    The term \emph{cryptocurrency} is often associated with concepts such as \emph{cryptoassets}, \emph{digital assets}, \emph{digital currency} and \emph{virtual currency}.

    \vspace{+2mm}
    \noindent\emph{Digital Asset.} The \SEC defines digital asset as an asset that is issued and/or transferred using \DLT, which includes but is not limited to virtual currencies\footnote{See \href{https://www.sec.gov/files/digital-assets-risk-alert.pdf}{The Division of Examinations’ Continued Focus on Digital Asset Securities}.}. Digital assets can be securities, currencies, properties, or commodities\footnote{ See \href{https://sgp.fas.org/crs/misc/R46208.pdf}{Digital Assets and SEC Regulation}.}.

    \vspace{+0.5mm}
    \noindent\emph{Cryptoasset.}  While no universally accepted definition of cryptoasset exists, they are broadly understood as digital representations of value or contractual rights, secured through cryptography\footnote{See section 2.4 of \href{https://www.fca.org.uk/publication/consultation/cp19-03.pdf}{CP19/3: Guidance on Cryptoassets}.}. Powered by \DLT, these assets can be electronically stored, transferred, or traded\footnote{See \href{https://eur-lex.europa.eu/eli/reg/2023/1114/oj}{Article 3 of the Markets in Crypto-assets (MiCA) Regulation}.}. In other words, a cryptoasset is a digital asset that uses cryptography to secure its functionality.

   \vspace{+0.5mm}
    \noindent\emph{Virtual Currency.} Virtual currency represents a digital form of value serving as a medium of exchange, a unit of account, and/or a store of value. It is a subset of digital assets without legal tender status in any jurisdiction~\cite{FATF2014}.

    \vspace{+0.5mm}
    \noindent\emph{Digital Currency.} Digital currency is a subset of digital assets with legal tender status. It is a tokenized, digital representation of a sovereign currency. It may be distributed by monetary authorities or entities backed by central bank money~\cite{ICU2019}.

    \vspace{+0.5mm}
    \noindent\emph{Cryptocurrency.} Cryptocurrency refers to a math-based, decentralized convertible virtual currency that is protected by cryptography~\cite{FATF2014}. It can be a subtype of digital currency or virtual currency, depending on its legal tender status.

    \vspace{+1mm}
    \noindent\textbf{Crypto Market Trends.} At the time of writing, there are over $8,000$ cryptocurrencies in circulation, boasting a combined market capitalization of approximately $1.6$t \USD and a daily trading volume of $63$b \USD\footnote{\url{https://coinmarketcap.com/charts/}, accessed on Jan~$20$,~$2024$.}. \BTC, \ETH and \USDT rank as the top three most actively traded cryptocurrencies. According to \href{https://www.statista.com/}{Statistia}, until Oct~$2023$, the global landscape features $1,492$ active exchanges, with a split of $62\%$ \CEXes (CEXs) and $38\%$ \DEXes. As of $2023$, the global cryptocurrency market has seen substantial growth, with the number of crypto users surpassing $420$m and the availability of $84$m crypto wallets worldwide. 
    The advent of cryptocurrency has unveiled various opportunities, such as trading and lending for both individuals and businesses, transcending traditional financial paradigms. It enhances financial market and payment system infrastructures, facilitating more efficient and secure transactions~\cite{us2022cryptoassets}. 

\section{Major Crypto Incidents and Regulatory Responses}\label{sec:incidents}

The cryptocurrency market, despite offering innovation and financial inclusion opportunities, has also witnessed numerous high-profile incidents. 
%


\vspace{+1mm}
\noindent\textbf{Mt. Gox Hack.} In $2014$, Mt. Gox, once the world's largest Bitcoin exchange, filed for bankruptcy following the theft of approximately $850,000$ \BTC, valued at around $450$m \USD at the time~\cite{trautman2014virtual,steele2020lessons}. This incident highlighted the security vulnerabilities within cryptocurrency exchanges and the need for enhanced regulatory oversight. In response, Japan, where Mt. Gox was based, enacted new legislation in $2017$ to regulate cryptocurrency exchanges. The \PSA\footnote{\href{https://www.japaneselawtranslation.go.jp/en/laws/view/3078/en}{Payment Services Act, Japan}.} was revised to require exchanges to implement stronger security measures, maintain adequate reserves, and undergo annual audits.

\noindent\textbf{Bitfinex Hack.} In $2016$, Bitfinex, a prominent cryptocurrency exchange, suffered a security breach that resulted in the loss of $120,000$ \BTC, worth about $72$m \USD then. This incident prompted a broader discussion on the security protocols of crypto exchanges and the necessity for regulatory intervention. Following the hack, the \CFTC fined Bitfinex for operating an unregistered exchange and for inadequate security measures\footnote{See \href{https://www.cftc.gov/PressRoom/PressReleases/7380-16}{CFTC Orders Bitcoin Exchange Bitfinex to Pay \$75,000 for Offering Illegal Off-Exchange Financed Retail Commodity Transactions and Failing to Register as a Futures Commission Merchant}.}, pushing for tighter security standards across the industry.

\noindent\textbf{FTX Collapse.} The collapse of FTX~\cite{jalan2023systemic,akyildirim2023understanding,fu2023ftx} in Nov~$2022$ marked one of the biggest failures in the industry's history. The downfall began when CoinDesk reported concerns about the financial health of Alameda Research, a trading firm closely tied to FTX, revealing that a large portion of its balance sheet was held in \FTT, the native token of FTX. This revelation led to a crisis of confidence among investors and users of the exchange, prompting a liquidity crunch as many attempted to withdraw their funds simultaneously. The core of FTX's collapse was attributed to a mix of factors, including the lack of clear separation between exchange operations and trading activities. The regulatory response to the FTX collapse was significant. The U.S. Department of Justice, the \SEC, and the Bahamas’ Financial Crimes Investigation Branch investigated the exchange and its executives for potential violations of financial regulations. 

\noindent\textbf{Terra-Luna Crash.} The collapse of Terra (\LUNA) and its stablecoin TerraUSD (\UST) in May $2022$ started when UST began to lose its peg~\cite{uhlig2022luna,liu2023anatomy,xiong2023leverage}. This destabilization was exacerbated by the interconnectedness of \UST with \LUNA through a mechanism that was supposed to maintain \UST's peg to the dollar by allowing the exchange of \UST for \LUNA and vice versa. Key factors contributing to the collapse included the high yield of $19.5\%$ offered by the \href{https://www.anchorprotocol.com/}{Anchor} to \UST depositors. The situation escalated when large withdrawals from Anchor preceded a bank run, with \UST depegging from the dollar and \LUNA's price collapsing, leading to a ``death spiral'' where \LUNA's supply inflated dramatically while its price plummeted to near zero. Regulatory responses to the crash have focused on scrutinizing stablecoins. The UK \HMT announced plans to regulate stablecoins as part of financial services legislation\footnote{See \href{https://www.gov.uk/government/publications/update-on-plans-for-the-regulation-of-fiat-backed-stablecoins}{Update on plans for the regulation of fiat-backed stablecoins}.}. The \FCA has also set out proposals for its regulation of fiat-backed stablecoins\footnote{See \href{https://www.fca.org.uk/publications/discussion-papers/dp23-4-regulating-cryptoassets-phase-1-stablecoins}{DP23/4: Regulating Cryptoassets Phase 1: Stablecoins}.}, indicating a shift towards more stringent oversight. 

\noindent\textbf{\TC Sanctions.}  \TC operates as a decentralized smart contract on the Ethereum blockchain, designed to enhance transaction privacy by mixing cryptocurrencies to obscure their origin, destination, and counterparties.  Despite its potential for legitimate use, TC has been criticized for facilitating money laundering by mixing the proceeds of cybercrimes. \TC was implicated in laundering over $7$b \USD worth of cryptocurrency since its inception in $2019$, including substantial amounts linked to North Korean state-sponsored hacking group Lazarus Group and other illicit activities such as the Harmony Bridge Heist and the Nomad Heist. The U.S. \OFAC sanctioned \TC on Aug~$8$,~$2022$\footnote{See \href{https://home.treasury.gov/news/press-releases/jy0916}{U.S. Treasury Sanctions Notorious Virtual Currency Mixer Tornado Cash}.}, marked a significant action against a blockchain mixer~\cite{wang2023blockchain}. The sanctions were implemented under Executive Order $13694$\footnote{See \href{https://www.govinfo.gov/app/details/CFR-2016-title3-vol1/CFR-2016-title3-vol1-eo13694}{3 CFR 13694 - Executive Order 13694 of April 1, 2015.}.}, aimed at combating the use of cryptocurrencies in illegal activities. 

\vspace{+1mm}
These significant incidents in the crypto market underscore the inherent risks and regulatory gaps. They have sparked significant regulatory actions, highlighting the urgent need for a more robust regulatory framework.

\section{Challenges in Cryptocurrency Regulation}\label{sec:chanllenges}

Regulating cryptocurrencies presents distinctive challenges due to their unique attributes and risks closely tied to the underlying blockchain technology. This section investigates the key challenges that demand regulatory attention.

\vspace{+1mm}
\noindent\textbf{C1: Blockchain Scalability vs Cryptocurrency Financial Inclusion.}
Cryptocurrencies hold significant potential to enhance financial inclusion, as emphasized by discussions in~\cite{pantin2023financial,ozili2022cbdc}. They offer alternative access to financial services for unbanked and underbanked populations, challenging traditional banking systems' exclusionary practices~\cite{pantin2023financial}. However, cryptocurrencies face challenges to widespread adoption. Despite the rising market capitalization of cryptocurrencies, their market value remains small compared to the traditional financial markets~\cite{policyIssue2020,policyIssue2023}. This discrepancy is largely attributed to blockchain scalability issues. For example, Bitcoin and Ethereum can handle 7 and 20-30 transactions per second, yet they encounter significant consensus delays~\cite{cong2018blockchain,zhou2020solutions,chauhan2018blockchain,xie2019survey,sanka2021systematic}. The limited transaction processing capacity and extended confirmation times directly impact cryptocurrency adoption rates and financial inclusion. Moreover, the modest adoption of cryptocurrency markets implies that regulatory motivation may be lessened. Regulators might perceive the costs of implementing and enforcing regulations on the cryptocurrency market as outweighing the benefits, given the market's current size and level of financial inclusion.

\vspace{+1mm}
\noindent\textbf{C2: Blockchain Pseudonymity vs User Privacy.} The pseudonymous feature of blockchain technology~\cite{li2019toward,politou2019blockchain}, while crucial for user privacy, significantly complicates the regulation of cryptocurrencies, especially in the context of combating fraudulent activities. Various countries have intensified efforts to tackle fraud and other illegal endeavors associated with cryptocurrencies, enacting specific laws and empowering regulatory bodies to oversee and mitigate such risks. For example, through agencies such as \SEC and \CFTC, the U.S. has implemented regulations targeting fraudulent activities in the cryptocurrency space. 

However, the inherent pseudonymity of blockchain transactions poses a substantial challenge to regulatory efforts~\cite{policyIssue2023,lianos2019regulating}. While transactions are recorded on a public ledger, they are linked to cryptographic addresses rather than personal identities. This level of anonymity complicates regulatory efforts to combat money laundering, fraud, and other illicit activities, as identifying the individuals behind transactions can be exceedingly difficult without infringing on privacy rights. The pseudonymity characteristic of blockchain not only hinders the identification and prosecution of fraudulent activities but also raises concerns about the balance between effective regulation and the protection of user privacy.

\vspace{+1mm}
\noindent\textbf{C3: Blockchain Decentralization vs Legal Accountability.} The decentralized nature of blockchain technology, a foundational aspect that drives the innovation and security behind cryptocurrencies, presents significant challenges for regulatory oversight and legal accountability. This decentralization means that unlike traditional financial systems, which have clear hierarchies and regulated entities, blockchain networks operate without a central authority. This structure complicates regulatory efforts to enforce compliance and accountability. For instance, in the case of fraudulent activities or disputes, pinpointing responsibility within a decentralized network can be exceedingly difficult. 

There is no single entity or group that regulators can hold accountable in the same way they would with traditional financial institutions. This challenge is starkly illustrated by the infamous \DAO hack~\cite{dhillon2021dao}, which occurred in Jun~$2016$. The \DAO was an innovative venture capital fund built on the Ethereum blockchain, designed to operate on smart contracts without any central authority. However, it became the target of a major exploit that led to the theft of approximately $3.6$m \ETH. The absence of a central authority complicated the legal response to the incident. The challenge for regulators is to develop mechanisms that can enforce legal accountability within these decentralized systems, ensuring that while blockchain's innovative and open nature is preserved, a framework is also in place to protect participants.

\vspace{+1mm}
\noindent\textbf{C4: Blockchain Cross-border Nature vs Jurisdictional Issues.} The cross-border nature of blockchain technology poses challenges to the legal regulation of cryptocurrencies across different countries. This global characteristic leads to a patchwork of regulatory approaches, with some nations opting for stringent controls while others adopt a more laissez-faire stance. 

Such disparities can result in regulatory arbitrages~\cite{fung2023regulatory}, which become evident as businesses and individuals seek to exploit these differences by basing their operations in jurisdictions with more favorable regulatory environments. For example, in response to the stringent regulations imposed by countries where \ICOs are heavily scrutinized, many projects have chosen to launch their \ICOs in jurisdictions with more lenient regulatory environments. Countries such as Switzerland and Singapore have emerged as popular destinations for \ICOs, thanks to their accommodating legal frameworks. Bores~\etal~\cite{Bores2023regulating} suggests that the costs associated with moving between different exchanges and legal territories are comparatively minimal. This strategic relocation allows projects to bypass the rigorous compliance requirements found in stricter jurisdictions, potentially leading to a regulatory race to the bottom. Therefore, the significance of international cooperation and the pursuit of harmonized regulation grows ever more critical.

\section{Cryptocurrency Regulation: Key Considerations}\label{sec:considerations}

In light of the challenges discussed earlier regarding cryptocurrency regulation, we present a series of questions that warrant the attention of regulators.

\vspace{+1mm}
\noindent\textbf{Q1: Should cryptocurrencies be regulated?} Regulators must weigh the advantages and disadvantages of regulating cryptocurrencies. Ensuring their inclusion within regulatory frameworks could enhance consumer protection and promote financial stability. Conversely, overregulation could hinder innovation and impede financial inclusion. Striking the right balance is crucial to harness the benefits of cryptocurrencies while mitigating potential risks.

\vspace{+1mm}
\noindent\textbf{Q2: Should regulators adapt existing frameworks or develop new ones?}  Adapting existing frameworks may facilitate regulation but could lead to ill-fitting rules. Conversely, creating bespoke regulations tailored to cryptocurrencies may ensure a more precise approach but could be time-consuming and resource-intensive. The choice impacts the effectiveness of regulatory efforts.

\vspace{+1mm}
\noindent\textbf{Q3: Who are the primary targets of cryptocurrency regulation?} Regulators must identify whether they focus on the technology, users, service providers, or specific activities within the cryptocurrency ecosystem.  This decision takes on heightened significance, particularly within the context of the decentralized crypto market, which operates without the presence of financial intermediaries.

\vspace{+1mm}
\noindent\textbf{Q4: What are the approaches for cryptocurrency regulation?} The quest for viable strategies in cryptocurrency regulation revolves around the overarching objective of regulators to attain ``\emph{same risk, same regulatory outcomes}''. This principle seeks to ensure that cryptocurrencies, bearing risks similar to traditional financial instruments, are subject to regulatory measures that correspond proportionately. In pursuit of this goal, regulators must devise strategies that align the regulatory framework for cryptocurrencies with their risk profiles and functions, striving for parity in regulatory standards and oversight between these digital assets and traditional financial instruments.

\section{Global Regulatory Landscape}\label{sec:landscape}

In this section, we first retrieve a list of countries (and regions) from 
\href{https://worldpopulationreview.com/countries/by-gdp}{World Population Review}. We then manually investigate the regulatory framework of each country by reviewing related laws, regulations, official guidelines from regulatory bodies, government announcements, and news reports. Finally, we develop a comprehensive taxonomy to classify and compare their regulatory stances, approaches, and measures towards cryptocurrencies.

\subsection{Classification of Regulatory Stances and Measures}
We categorize the regulatory stance of a given country into one of the following categories: \emph{(i) General Ban}: All activities related to cryptocurrencies are prohibited; \emph{(ii) Partial Ban}: Some activities related to cryptocurrencies are prohibited; \emph{(iii) Restrictive Regulation}: Cryptocurrency-related activities are subject to stringent regulations to mitigate potential risks; \emph{(iv) Supportive Regulation}: The country regulates cryptocurrency-related activities while fostering the development of the crypto market; \emph{(v) Concerned}: There is no specific regulation on cryptocurrencies, but there are expressed concerns regarding potential risks; \emph{(vi) Laissez-faire}: Cryptocurrencies remain unregulated; \emph{(vii) Legal Tender}: Cryptocurrencies are recognized as legal tender; \emph{(viii) No information}: There is no available information concerning the regulatory stance of the given country.

If a country has implemented regulation for cryptocurrencies, we further categorize its regulatory approach into one of the following three categories: \emph{(i) Existing Framework}: The country integrates cryptocurrency regulations within its existing legal and regulatory framework; \emph{(ii) New Framework}: The country establishes a new, bespoke regulatory framework specifically designed for cryptocurrencies; \emph{(iii) Hybrid Approach}: The country employs a combined strategy, leveraging both existing legal structures and introducing new regulations.

We further explore the regulatory measures adopted by countries worldwide, including \AML/\CTF, \ICOs, exchanges, taxation, stablecoin, \NFTs, and \DeFi. For each measure, we classify a country's regulatory status into one of three categories: \emph{(i) Regulated}, \emph{(ii) Unregulated}, \emph{(iii) Not Applicable}, \emph{(iv) Unclear}.

\subsection{Overview of the Global Regulatory Landscape}

    \begin{figure}[t]
    \centering
    \includegraphics[width=\columnwidth]{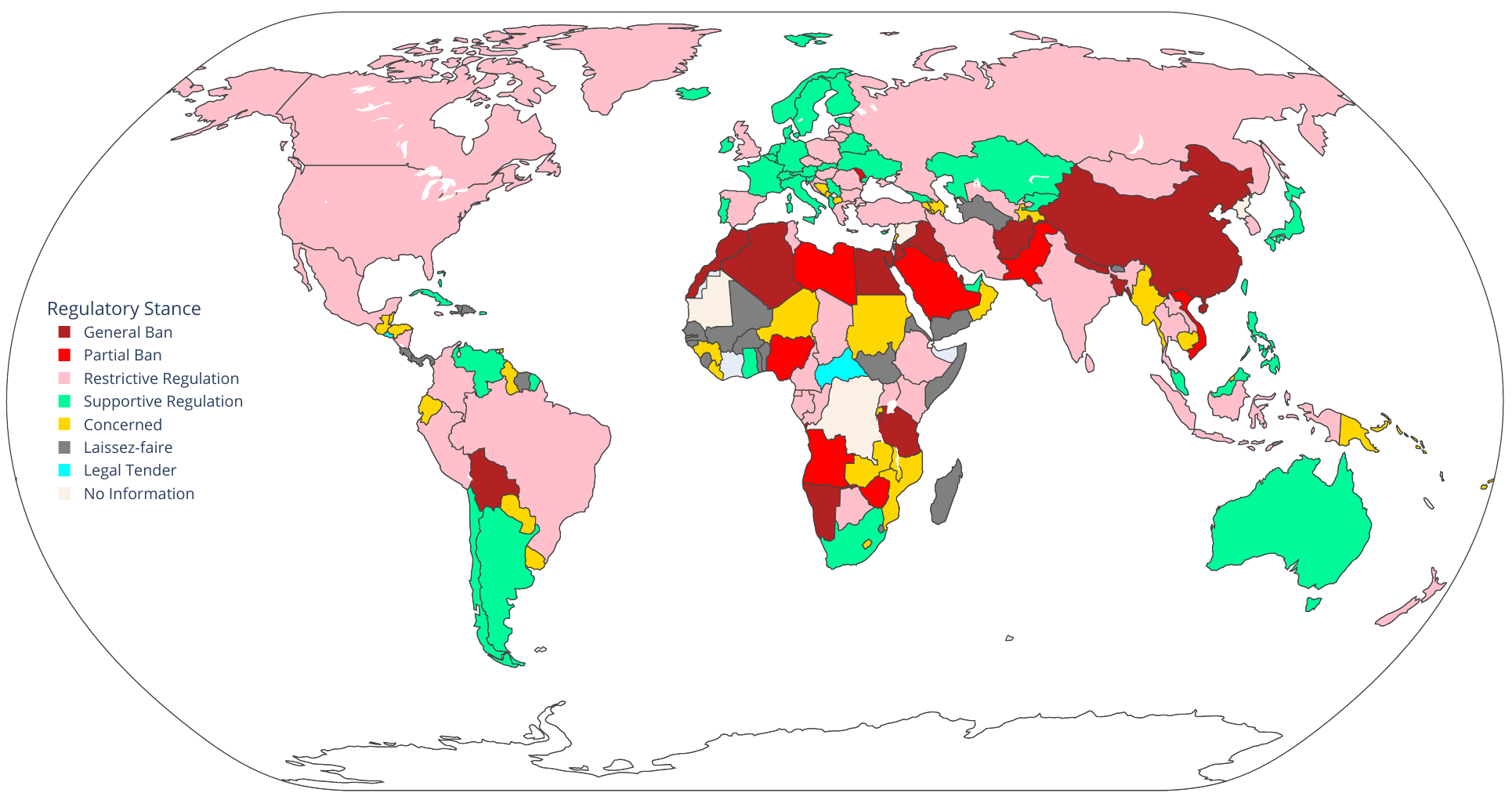}
    \caption{Overview of the Global Cryptocurrency Regulatory Landscape.}
    \label{fig: reg_world}
    \end{figure}

We provide a visualization of our classification of regulatory stances in Figure~\ref{fig: reg_world}. It is evident that there is a substantial variation in the regulatory attitudes towards cryptocurrencies across different countries. We observe that 58 (27.4\%) countries have adopted a supportive stance towards cryptocurrency regulation, indicating a strong desire to foster innovation. Four (1.9\%) countries, such as El Salvador, have even recognized cryptocurrencies as legal tender, suggesting a growing recognition of the potential benefits and opportunities presented by cryptocurrencies. Nevertheless, 14 (6.6\%) countries have implemented a general ban on cryptocurrencies, while nine (4.2\%) countries have enacted partial bans. This reflects profound concerns regarding the risks cryptocurrencies pose, including their potential to facilitate financial crimes, disrupt monetary policy, and compromise economic stability. 46 (21.7\%) countries have adopted restrictive regulations, aiming to mitigate potential risks and ensure a secure financial environment. These measures include stringent compliance requirements and specific legal frameworks designed to protect consumers, prevent fraud, and combat money laundering and terrorism financing.  

Interestingly, we note that cryptocurrency remains unregulated in at least 71 countries. Specifically, 37 (17.5\%) countries have merely expressed concerns about the risks associated with cryptocurrencies, while the remaining 34 (16.0\%) have adopted a laissez-faire approach, choosing not to intervene or regulate the crypto sector. This hands-off approach can lead to a lack of consumer protection, where investors may face heightened risks without any safeguarding measures or legal recourse in case of fraud or market manipulation. Furthermore, the absence of regulation can lead to regulatory arbitrage, where entities exploit the discrepancy in regulatory landscapes across different jurisdictions to bypass stringent regulations by operating in or transacting with entities in less regulated countries,  potentially undermining the efforts of more regulated nations. 

We further analyze the regulatory approaches of countries that implement cryptocurrency regulations. We observe that 20 of these countries have integrated cryptocurrency regulation within their existing legal frameworks, adapting traditional financial laws to encompass the unique characteristics of cryptocurrencies. Meanwhile, 40 countries have developed bespoke frameworks specifically tailored for the regulation of cryptocurrencies, reflecting a recognition of the need for specialized rules to address the novel challenges and opportunities they present. Additionally, 48 countries have adopted hybrid approaches, suggesting a balanced effort to ensure regulatory efficiency.

\begin{table}[tbh]
    \centering
    \resizebox{\textwidth}{!}{
    \begin{tabular}{c|c:c:c:c:c:c:c}
    \toprule
     Regulatory Measure & AML/CFT & Taxation & 
     Registration & ICO & StableCoin & NFT & DeFi\\
     \midrule
     \# Participating Countries & 102& 87 & 101 &	74 &39	&0	&0 \\
     \% Participating Countries & 48.1\% & 41.0\%	 &47.6\%	& 34.9\%& 18.4\% & 0\%  & 0\%		 \\
     \bottomrule
    \end{tabular}
    }
    \caption{Global Adoption of Regulatory Measures.}
    \label{tab:measure}
\end{table}

We then investigate the specific regulatory measures enacted by various countries. Table~\ref{tab:measure} shows that reveals that 48.1\% of the countries are actively taking steps to incorporate cryptocurrencies into their AML/CTF regime. In addition, 41\% of countries have developed tax frameworks specifically for cryptocurrencies. We also discover that 47.6\% of countries have implemented the registration requirement for crypto service providers and  34.9\% of countries have established regulations for \ICOs. However, only 18.4\% of countries have established regulations specifically targeting stablecoins. Moreover, although countries such as the UK and U.S. have shown interest in incorporating NFTs and DeFi into their regulatory frameworks, no nation has established regulations specifically tailored to \NFTs or \DeFi. This regulatory gap may attract illicit activities, such as wash trading~\cite{victor2021detecting,von2022nft} and money laundering~\cite{wu2023towards,mosna2023nfts} in the \NFT and \DeFi domains.

\section{Case Study}\label{sec:case_study}
Section~\ref{sec:landscape} provides an overview of the regulatory landscape across all countries. To better understand the variation in regulatory progress worldwide, this section proceeds to study the regulatory landscape within selected economies. 

\subsection{United States}

In the U.S., legislative bodies and regulators have yet to establish a framework tailored to cryptocurrencies that directly equates to the traditional regulation seen in securities or commodities markets.  They mainly apply existing financial regulations to cryptocurrencies. Cryptoassets may be categorized as securities or commodities based on their functionalities and the context of their use. 

The \SEA regulates cryptocurrencies by evaluating whether they meet the definition of ``security'' or ``investment contract''\footnote{See \href{https://www.govinfo.gov/content/pkg/COMPS-1885/pdf/COMPS-1885.pdf}{Section 3 of the Securities Exchange Act of 1934}.}, particularly through the application of the Howey Test\footnote{See \href{https://www.sec.gov/corpfin/framework-investment-contract-analysis-digital-assets}{Framework for `Investment Contract' Analysis of Digital Assets}.}. If a cryptocurrency is deemed to offer an investment opportunity where profits are expected to be derived from the efforts of others, it may be considered a security and thus fall under the \SEC's jurisdiction. In fact, the SEC has initiated enforcement actions against cryptocurrency issuers for not registering their cryptocurrencies as securities, emphasizing regulatory compliance. In Dec~$2020$, the SEC sued Ripple Labs\footnote{See \href{https://www.sec.gov/news/press-release/2020-338}{SEC Charges Ripple and Two Executives with Conducting 1.3 Billion Unregistered Securities Offering}.}, claiming Ripple raised over $1.3$b \USD via unregistered securities offerings through \XRP sales starting in 2013. Ripple contended that \XRP was a cryptocurrency and did not need securities registration. However, the SEC has taken a firm stance on \XRP, categorizing it as a security and not a currency.  This case highlights regulatory efforts to bring cryptoassets under traditional securities law.

Under the \CEA,the \CFTC regulates cryptocurrencies deemed as ``commodities'', focusing on futures, swaps, and derivative contracts tied to digital assets. The CFTC's jurisdiction encompasses the oversight of trading platforms, ensuring market integrity, preventing manipulation, and protecting investors in commodity-linked cryptocurrency markets. The \CFTC has actively pursued legal actions and regulatory oversight against several cryptocurrency firms. In Sep~$2015$, the CFTC engaged in regulatory action against Coinflip\footnote{See \href{https://www.cftc.gov/PressRoom/SpeechesTestimony/opamcginley1}{The CFTC’s Actions in the Derivatives Markets for Digital Assets}.}, the entity behind the Derivabit platform, which facilitated trading in Bitcoin options and futures. This regulatory step reinforced the CFTC's position on considering Bitcoin and other similar digital currencies as ``commodities''. Aitan Goelman, the \CFTC’s Director of Enforcement, commented that ``While there is a lot of excitement surrounding Bitcoin and other virtual currencies, innovation does not excuse those acting in this space from following the same rules applicable to all participants in the commodity derivatives markets''\footnote{See \href{https://www.cftc.gov/PressRoom/PressReleases/7231-15}{CFTC Orders Bitcoin Options Trading Platform Operator and its CEO to Cease Illegally Offering Bitcoin Options and to Cease Operating a Facility for Trading or Processing of Swaps without Registering}.}. In Oct~$2020$, the \CFTC took action against BitMEX\footnote{See \href{https://www.cftc.gov/PressRoom/SpeechesTestimony/opamcginley1}{The CFTC’s Actions in the Derivatives Markets for Digital Assets}.}, a prominent cryptocurrency exchange, accusing it of operating an unregistered trading platform. The allegations included failing to implement required \AML measures.

Overall, the regulatory actions and disputes between the \SEC, \CFTC, and cryptocurrency companies imply regulatory ambiguities surrounding the classification of cryptocurrencies. These cases underscore the lack of clear regulatory guidelines for cryptocurrencies, leading to debates on whether they should be treated as securities, commodities, or a new asset class. These disputes underscore the need for clearer regulatory guidelines to distinguish between different types of cryptocurrencies and their corresponding regulatory obligations.

\subsection{European Union}

The EU initially approached the regulation of cryptoassets through a series of directives and regulations aimed at integrating these cryptoassets within the existing regulatory frameworks. The \MiFiD mainly regulates cryptoassets considered ``financial instruments'' (e.g., transferable securities)\footnote{See \href{https://eur-lex.europa.eu/legal-content/EN/TXT/?uri=celex\%3A32014L0065}{Section C of Annex I, MiFiD II}.}. The amended \AMLD  introduces the \AML and \CTF regimes for cryptoasset providers, including the custodial wallet providers and providers engaged in exchange services between virtual currencies and fiat currencies. Certain types of cryptoasset may also fall within \EMD II if qualified as e-money, or \PSD if used as a medium for executing payment services. Nevertheless, since individual Member States can interpret derivatives at their discretion, the regulation remains unharmonized, leading to legal uncertainties across the EU~\cite{van2023markets}. Moreover, the existing regulatory framework captures only a subset of cryptoassets. This left investors vulnerable to risks, especially in areas outside the scope of consumer protection laws.

To address such regulatory gaps, in May~$2023$, the \MiCA regulation\footnote{\href{https://eur-lex.europa.eu/eli/reg/2023/1114/oj}{Regulation (EU) 2023/1114 of the European Parliament and of the Council of 31 May 2023 on markets in crypto-assets}.} was enacted. \MiCA introduces a harmonized regulatory framework for cryptoasset markets across the EU. \MiCA covers cryptoassets that are not regulated by existing financial services legislation\footnote{For example, \MiCA does not cover security tokens, which would qualify as transferable securities under \MiFiD. }. For example, it covers three types of cryptoassets: \emph{(i)} Asset-reference tokens that stabilize their value by referencing the values of one or more assets; \emph{(ii)} E-money tokens that stabilize their value by referencing the value of a single fiat currency; \emph{(iii)} Other crypto-assets (e.g., utility tokens). \MiCA applies to cryptoasset issuers and  \CASPs, who must acquire authorizations from the \NCA in the respective member states. \MiCA covers various activities, such as the issuance, offering to the public, and admission to trading of cryptoassets. Along with \MiCA, the \TFR, published in Jun~$2023$, introduced new rules on the information on originators and beneficiaries accompanying transfers of cryptoassets\footnote{\href{https://eur-lex.europa.eu/legal-content/EN/TXT/?uri=CELEX\%3A32023R1113}{Regulation (EU) 2023/1113 of the European Parliament and of the Council of 31 May 2023 on information accompanying transfers of funds and certain crypto-assets}.}.

To summarize, the EU initially approached cryptoassets with caution, integrating them into existing financial frameworks to manage risks. Over time, recognizing the need for a more tailored approach, the EU shifted towards establishing a unified and harmonized regulatory environment.

\subsection{United Kingdom} \label{case:uk}

The UK's approach to cryptoasset regulation involves the incorporation of existing regulatory frameworks. In 2018, the ``Cryptoassets Taskforce (CATF)'' was formed, comprising the \FCA, the \BoE, and \HMT\footnote{See \href{https://assets.publishing.service.gov.uk/media/5bd6d6f0e5274a6e11247059/cryptoassets_taskforce_final_report_final_web.pdf}{Cryptoassets Taskforce}.}. The CATF notes that the crypto market presents both new opportunities and potential challenges. It recommends that decisive measures be implemented to mitigate the risks of cryptoassets that fall within existing regulatory frameworks.  In Feb~$2023$, the \HMT released a consultation paper entitled ``Future Financial Services Regulatory Regime for Cryptoassets''. This document outlines the government's plan to integrate the regulation of cryptoassets into the existing \FSMA framework\footnote{See section 2.1 of \href{https://assets.publishing.service.gov.uk/media/653bd1a180884d0013f71cca/Future_financial_services_regulatory_regime_for_cryptoassets_RESPONSE.pdf}{Future financial services regulatory regime for cryptoassets: Response to the consultation and call for evidence}.}. The government adopts ``technology agnostic'' approaches and values the principle of ``same risks, same regulatory outcomes'' in terms of regulating the crypto market and the traditional financial market\footnote{See section 1.12 of \href{https://assets.publishing.service.gov.uk/media/653bd1a180884d0013f71cca/Future_financial_services_regulatory_regime_for_cryptoassets_RESPONSE.pdf}{Future financial services regulatory regime for cryptoassets: Consultation and call for evidence}.}. Hence, developing a standalone bespoke regime outside of the \FSMA framework would risk creating an uneven playing field between cryptoasset companies and traditional financial institutions. As shown in Table~\ref{tab:uk_reg_overview}, the existing framework applicable to cryptoassets regulation mainly encompasses \emph{(i)} \FSMA 2023\footnote{ \href{https://www.legislation.gov.uk/ukpga/2023/29/contents}{Financial Services and Markets Act 2023}.} (including \RAO 2001\footnote{\href{https://www.legislation.gov.uk/uksi/2001/544/contents}{The FSMA 2000 (Regulated Activities) Order 2001}.} and \FPO 2023\footnote{\href{https://www.legislation.gov.uk/ukdsi/2023/9780348246490/body}{The FSMA 2000 (Financial Promotion) (Amendment) Order 2023}.}), \emph{(ii)} the \AML/\CTF regulations 2017\footnote{\href{https://www.legislation.gov.uk/uksi/2017/692/contents/made}{The Money Laundering, Terrorist Financing and Transfer of Funds (Information on the Payer) Regulations 2017}.}, \emph{(iii)} the \PSRs 2017, and \emph{(iv)} \EMRs 2011. In fact, the \FSMA contains several provisions that amend the existing regimes to explicitly accommodate the regulation of cryptoassets. For example, amendments to the \FSMA allow for the explicit inclusion of cryptoassets within the \RAO and \FPO regime\footnote{See section 69 of FSMA 2023.}

The Consultation defines four types of cryptoassets, including security tokens, exchanges tokens, utility tokens, and \NFTs (See Appendix~\ref{appendix_uk_taxonomy}). Some of these tokens fall within the existing regulatory framework. For example, security tokens fall within the regulatory framework, as they provide rights and obligations akin to ``specified investments''\footnote{See \href{https://www.legislation.gov.uk/uksi/2001/544/part/III/made}{Part III of RAO}.} as set out in the \RAO. While exchange tokens and utility tokens generally remain unregulated, if they meet the definition of e-money, they will fall within the scope of \EMRs\footnote{See section 3.7 of \href{https://www.fca.org.uk/publication/policy/ps19-22.pdf}{ PS19/22: Guidance on Cryptoassets}.}. Furthermore, these tokens are subject to the \PSRs if used to facilitate regulated payment services\footnote{See section 3.54 of \href{https://www.fca.org.uk/publication/consultation/cp19-03.pdf}{CP19/3: Guidance on Cryptoassets}.}.

In fact, the \HMT pursues a ``phased approach'' to regulating cryptoassets. In Phase 1, the government intends to regulate fiat-backed stablecoins.   Under the \RAO regime, UK-based stablecoins issuers and custodians should seek authorizations from the \FCA\footnote{See section 2.24 of \href{https://www.fca.org.uk/publication/discussion/dp23-4.pdf}{DP23/4: Regulating Cryptoassets Phase 1: Stablecoins}.}. The payment transactions in relation to fiat-backed stablecoins by firms in the UK will be brought within the scope of the \PSRs\footnote{See section 2.4 of \href{https://assets.publishing.service.gov.uk/media/653a82b7e6c968000daa9bdd/Update_on_Plans_for_Regulation_of_Fiat-backed_Stablecoins_13.10.23_FINAL.pdf}{Update on Plans for the Regulation of Fiat-backed Stablecoins}.}. In addition, FSMA 2023 also introduces digital settlement assets to address payment stablecoin\footnote{See section 86 of \href{https://www.legislation.gov.uk/ukpga/2023/29/contents}{FSMA 2023}.}. In Phase 2, the regulation will be extended to other categories of cryptoassets. Broader cryptoasset activities, such as issuance, lending and borrowing, leverage, and exchange activities, will be brought into the regulatory perimeter\footnote{See section 1.16 of \href{https://assets.publishing.service.gov.uk/media/653bd1a180884d0013f71cca/Future_financial_services_regulatory_regime_for_cryptoassets_RESPONSE.pdf}{Future financial services regulatory regime for cryptoassets: Response to the consultation and call for evidence}.}. Cryptoasset businesses registered under the \AML/\CTF regime will not receive automatic authorization for cryptoasset activities during Phase 1 or 2. Such firms need to obtain authorization under \FSMA or \PSRs\footnote{See section 3.9 of \href{https://assets.publishing.service.gov.uk/media/653bd1a180884d0013f71cca/Future_financial_services_regulatory_regime_for_cryptoassets_RESPONSE.pdf}{Future financial services regulatory regime for cryptoassets: Response to the consultation and call for evidence}.}.

To summarize, the UK's regulatory approach towards cryptoassets has evolved to be both inclusive and prudent, aiming to adapt its existing legal frameworks to the unique facets of cryptoassets. The phased approach allows for gradual integration into the financial landscape, maintaining a balance between fostering innovation and ensuring consumer protection and market stability.

\subsection{China}

China's approach to regulating cryptocurrencies has been characterized by a series of stringent measures aimed at controlling the associated risks. 

In $2013$, the \PBC, along with five other departments, issued the ``Notice on Preventing Bitcoin Risk'', the first official document to address the legal status of cryptocurrencies in China\footnote{See \href{https://www.gov.cn/gzdt/2013-12/05/content_2542751.htm}{Notice on Preventing Bitcoin Risk}.}. It declared that Bitcoin does not hold the same legal status as a currency and prohibited financial and payment institutions from engaging in Bitcoin-related transactions. This notice aimed to protect the yuan's status and prevent money laundering.

The rapid rise of \ICOs as a method for funding cryptocurrency projects prompted the Chinese government to take further action. In 2017, the \PBC, along with seven other departments, released an announcement that outright banned \ICOs in China. The ``Announcement on Preventing Risks of Token Issuance Financing'' labeled \ICOs as an unauthorized fundraising tool that may involve financial fraud, illegal issuance of securities, and other criminal activities\footnote{See \href{https://www.gov.cn/xinwen/2017-09/04/content_5222657.htm}{Announcement on Preventing Risks of Token Issuance Financing}.}. This move was aimed at protecting investors from risky ventures. 

In response to the resurgence of speculative trading in cryptocurrencies, in 2021, the \PBC and ten other departments issued a more comprehensive notice. The ``Notice on Further Preventing and Handling the Risks of Speculation in Virtual Currency Transactions'' reinforced the ban on cryptocurrency trading\footnote{See \href{https://www.gov.cn/zhengce/zhengceku/2021-10/08/content_5641404.htm}{Notice on Further Preventing and Handling the Risks of Speculation in Virtual Currency Transactions}.}. This document highlighted the government's stance on cryptocurrencies as not being recognized as legal tender and reiterated the prohibition of their use. The notice also targeted cryptocurrency mining, declaring it an illegal activity.

Through these key policy documents, it is evident that the Chinese government maintains a strict and evolving stance on cryptocurrency regulation. From the initial measures to prevent Bitcoin risks to the comprehensive prohibition of \ICOs and further banning cryptocurrency transactions and mining activities, China's regulatory policies have become progressively clear and strengthened.

\subsection{India}

India's regulatory stance can be described as a complex and evolving trajectory, marked by periods of skepticism, regulatory ambiguity, and gradual acceptance. 

The initial phase of India's approach towards cryptocurrencies can be traced back to $2013$, when the \RBI issued its first cautionary advice\footnote{See \href{https://www.rbi.org.in/commonman/English/Scripts/PressReleases.aspx?Id=2152}{RBI cautions users of Virtual Currencies}.} to the Indian public about the risks associated with trading in Bitcoin and other digital currencies. This period was marked by a significant degree of uncertainty, as the RBI highlighted concerns related to volatility and the potential use of cryptocurrencies for illicit activities. The absence of any formal regulatory framework left the market in a state of limbo, with investors and operators navigating a grey area of legal and financial risks.

The situation took a more definitive turn in Apr~$2018$, when the \RBI directed all regulated financial institutions to cease dealing with individuals or businesses transacting in cryptocurrencies, effectively cutting off banking services to the crypto sector\footnote{See \href{https://www.rbi.org.in/commonman/Upload/English/PressRelease/PDFs/PR264205042018.PDF}{Statement on Developmental and Regulatory Policies}.}. However, in Mar~$2020$, the Supreme Court overturned the \RBI's ban, citing the disproportionate nature of the regulatory response\footnote{See \href{https://www.reuters.com/article/idUSKBN20R0KW/}{India's top court strikes down RBI banking ban on cryptocurrency}.}. This decision was hailed as a significant victory for the crypto sector in India, signaling a potential shift towards a more accommodating regulatory environment.

The contradictory stances of different Indian regulatory bodies lead to a complex and uncertain environment for investors and businesses in the crypto space. This inconsistency has not only impacted the operational realities of crypto businesses but has also influenced investor sentiment and market stability. The ongoing uncertainty necessitates a balanced and clear regulatory framework.

\section{Discussion}\label{sec:discussion}

Through the analysis of cryptocurrency regulations across different jurisdictions, we observe significant variations. Some have outright banned cryptocurrency activities, while others supported innovation. Some have applied existing regulations, whereas others established bespoke frameworks. This finding emphasizes the urgent need for international cooperation and harmonized regulation.

\vspace{+0.5mm}
India's regulatory journey underscores the need for consistency and collaboration among regulatory agencies to avoid conflicting directives and ensure cohesive policy implementation. From the U.S. experience, the importance of clear legal definitions and the establishment of distinct responsibilities for regulatory bodies emerge as critical for avoiding jurisdictional overlaps and ensuring effective regulation. The UK's gradual, phased approach to crypto regulation highlights the benefits of allowing the market and stakeholders to adjust over time, avoiding sudden disruptions. The EU's exploration into a new regulatory framework for cryptocurrencies illustrates the attempt to harmonize crypto regulation with existing financial systems. In essence, cryptocurrency regulation should be tailored to reflect the country's unique circumstances and market dynamics.

In addition, our study identifies a notable absence of explicit regulation for \NFTs worldwide, a concerning observation given their unique attributes and associated risks. The lack of regulatory oversight could inadvertently facilitate illegal activities~\cite{von2022nft}. Therefore, we argue that the regulation of \NFTs should not be overlooked. Regulators across the globe should actively engage in devising effective regulatory measures to regulate the \NFT market. 

Furthermore, we discover that \DeFi remains largely unregulated. For instance, \MiCA states that ``Where crypto-asset services are provided in a fully decentralized manner without any intermediary, they should not fall within the scope of this Regulation''. \MiCA appears to exclude \DeFi from its regulation, yet the term ``fully decentralized'' is exceedingly ambiguous, leading to the question of how decentralized a platform must be to qualify as fully decentralized? This vague regulatory stance may stem from the inherent complexities of regulating DeFi. However, considering the unique risks and characteristics of \DeFi, we believe that regulators should adopt a clearer stance towards its regulation.

\section{Conclusion}\label{sec:conclusion}

This paper explores the global regulatory landscape for cryptocurrencies. Through classification and analysis of the regulatory stances, approaches, and measures implemented by various countries, we highlight the diversity in global regulatory developments. This underscores the importance of international cooperation and the pursuit of harmonized regulations. In addition, we discover that cryptocurrencies still remain unregulated in at least 71 countries at the time of writing. Moreover, we observe that no jurisdictions have established specific regulatory frameworks for \NFTs or \DeFi. Our findings indicate that further research and policy attention are needed to address such regulatory gaps. We hope this paper can provide practical insights that contribute to a better understanding of the global trends in cryptocurrency regulation.

%
%
\bibliographystyle{ieeetr}
\bibliography{references}

\begin{thebibliography}{10}

\bibitem{blandin2019global}
A.~Blandin, A.~S. Cloots, H.~Hussain, M.~Rauchs, R.~Saleuddin, J.~G. Allen,
  B.~Z. Zhang, and K.~Cloud, ``Global cryptoasset regulatory landscape study,''
  {\em University of Cambridge Faculty of Law Research Paper}, no.~23, 2019.

\bibitem{cumming2019regulation}
D.~J. Cumming, S.~Johan, and A.~Pant, ``Regulation of the crypto-economy:
  Managing risks, challenges, and regulatory uncertainty,'' {\em Journal of
  Risk and Financial Management}, vol.~12, no.~3, p.~126, 2019.

\bibitem{mezquita2023cryptocurrencies}
Y.~Mezquita, D.~P{\'e}rez, A.~Gonz{\'a}lez-Briones, and J.~Prieto,
  ``Cryptocurrencies, survey on legal frameworks and regulation around the
  world,'' in {\em International Congress on Blockchain and Applications},
  pp.~58--66, Springer, 2023.

\bibitem{werner2022sok}
S.~Werner, D.~Perez, L.~Gudgeon, A.~Klages-Mundt, D.~Harz, and W.~Knottenbelt,
  ``Sok: Decentralized finance (defi),'' in {\em Proceedings of the 4th ACM
  Conference on Advances in Financial Technologies}, pp.~30--46, 2022.

\bibitem{FATF2014}
{FATF}, ``Virtual currencies: Key definitions and potential aml/cft risks,''
  2014.

\bibitem{ICU2019}
{International Communication Union}, ``Taxonomy and definition of terms for
  digital fiat currency,'' 2019.

\bibitem{us2022cryptoassets}
U.~D. of~the Treasury, ``Crypto-assets: Implications for consumers, investors,
  and businesses,'' 2022.

\bibitem{trautman2014virtual}
L.~J. Trautman, ``Virtual currencies; bitcoin \& what now after liberty
  reserve, silk road, and mt. gox?,'' {\em Richmond Journal of Law and
  Technology}, vol.~20, no.~4, 2014.

\bibitem{steele2020lessons}
S.~Steele and T.~Morishita, ``Lessons from mt gox: practical considerations for
  a virtual currency insolvency,'' in {\em Research Handbook on Asian Financial
  Law}, pp.~479--498, Edward Elgar Publishing, 2020.

\bibitem{jalan2023systemic}
A.~Jalan and R.~Matkovskyy, ``Systemic risks in the cryptocurrency market:
  Evidence from the ftx collapse,'' {\em Finance Research Letters}, vol.~53,
  p.~103670, 2023.

\bibitem{akyildirim2023understanding}
E.~Akyildirim, T.~Conlon, S.~Corbet, and J.~W. Goodell, ``Understanding the ftx
  exchange collapse: A dynamic connectedness approach,'' {\em Finance Research
  Letters}, vol.~53, p.~103643, 2023.

\bibitem{fu2023ftx}
S.~Fu, Q.~Wang, J.~Yu, and S.~Chen, ``Ftx collapse: a ponzi story,'' in {\em
  International Conference on Financial Cryptography and Data Security},
  pp.~208--215, Springer, 2023.

\bibitem{uhlig2022luna}
H.~Uhlig, ``A luna-tic stablecoin crash,'' tech. rep., National Bureau of
  Economic Research, 2022.

\bibitem{liu2023anatomy}
J.~Liu, I.~Makarov, and A.~Schoar, ``Anatomy of a run: The terra luna crash,''
  tech. rep., National Bureau of Economic Research, 2023.

\bibitem{xiong2023leverage}
X.~Xiong, Z.~Wang, X.~Chen, W.~Knottenbelt, and M.~Huth, ``Leverage staking
  with liquid staking derivatives (lsds): Opportunities and risks,'' {\em arXiv
  preprint arXiv:2401.08610}, 2023.

\bibitem{wang2023blockchain}
Z.~Wang, X.~Xiong, and W.~J. Knottenbelt, ``Blockchain transaction
  censorship:(in) secure and (in) efficient?,'' {\em Cryptology ePrint
  Archive}, 2023.

\bibitem{pantin2023financial}
L.~P. Pantin, ``Financial inclusion, cryptocurrency, and afrofuturism,'' {\em
  Northwestern University Law Review}, vol.~118, no.~3, pp.~621--690, 2023.

\bibitem{ozili2022cbdc}
P.~K. Ozili, ``Cbdc, fintech and cryptocurrency for financial inclusion and
  financial stability,'' {\em Digital Policy, Regulation and Governance},
  vol.~25, no.~1, pp.~40--57, 2022.

\bibitem{policyIssue2020}
D.~W. Perkins, ``Cryptocurrency: The economics of money and selected policy
  issues,'' {\em Congressional Research Service}, pp.~1--27, 2020.

\bibitem{policyIssue2023}
P.~Tierno, ``Cryptocurrency: The economics of money and selected policy
  issues,'' {\em Congressional Research Service}, pp.~1--27, 2020.

\bibitem{cong2018blockchain}
K.~Cong, Z.~Ren, and J.~Pouwelse, ``A blockchain consensus protocol with
  horizontal scalability,'' in {\em 2018 IFIP Networking Conference (IFIP
  Networking) and Workshops}, pp.~1--9, IEEE, 2018.

\bibitem{zhou2020solutions}
Q.~Zhou, H.~Huang, Z.~Zheng, and J.~Bian, ``Solutions to scalability of
  blockchain: A survey,'' {\em Ieee Access}, vol.~8, pp.~16440--16455, 2020.

\bibitem{chauhan2018blockchain}
A.~Chauhan, O.~P. Malviya, M.~Verma, and T.~S. Mor, ``Blockchain and
  scalability,'' in {\em 2018 IEEE international conference on software
  quality, reliability and security companion (QRS-C)}, pp.~122--128, IEEE,
  2018.

\bibitem{xie2019survey}
J.~Xie, F.~R. Yu, T.~Huang, R.~Xie, J.~Liu, and Y.~Liu, ``A survey on the
  scalability of blockchain systems,'' {\em IEEE network}, vol.~33, no.~5,
  pp.~166--173, 2019.

\bibitem{sanka2021systematic}
A.~I. Sanka and R.~C. Cheung, ``A systematic review of blockchain scalability:
  Issues, solutions, analysis and future research,'' {\em Journal of Network
  and Computer Applications}, vol.~195, p.~103232, 2021.

\bibitem{li2019toward}
Y.~Li, W.~Susilo, G.~Yang, Y.~Yu, X.~Du, D.~Liu, and N.~Guizani, ``Toward
  privacy and regulation in blockchain-based cryptocurrencies,'' {\em IEEE
  Network}, vol.~33, no.~5, pp.~111--117, 2019.

\bibitem{politou2019blockchain}
E.~Politou, F.~Casino, E.~Alepis, and C.~Patsakis, ``Blockchain mutability:
  Challenges and proposed solutions,'' {\em IEEE Transactions on Emerging
  Topics in Computing}, vol.~9, no.~4, pp.~1972--1986, 2019.

\bibitem{lianos2019regulating}
I.~Lianos, P.~Hacker, S.~Eich, and G.~Dimitropoulos, {\em Regulating
  blockchain: techno-social and legal challenges}.
\newblock Oxford University Press, 2019.

\bibitem{dhillon2021dao}
V.~Dhillon, D.~Metcalf, M.~Hooper, V.~Dhillon, D.~Metcalf, and M.~Hooper, ``The
  dao hacked,'' {\em Blockchain Enabled Applications: Understand the Blockchain
  Ecosystem and How to Make it Work for You}, pp.~113--128, 2021.

\bibitem{fung2023regulatory}
S.~Fung, K.~Obaid, and L.~Tam, ``Regulatory arbitrage and initial coin
  offerings,'' {\em Available at SSRN 4629540}, 2023.

\bibitem{Bores2023regulating}
R.~Bores and A.-M. Bores, ``Egulating cryptographic financial instruments from
  intent to execution,'' {\em uropean Journal of Accounting, Finance \&
  Business}, vol.~11, p.~1, 2023.

\bibitem{victor2021detecting}
F.~Victor and A.~M. Weintraud, ``Detecting and quantifying wash trading on
  decentralized cryptocurrency exchanges,'' in {\em Proceedings of the Web
  Conference 2021}, pp.~23--32, 2021.

\bibitem{von2022nft}
V.~von Wachter, J.~R. Jensen, F.~Regner, and O.~Ross, ``Nft wash trading:
  Quantifying suspicious behaviour in nft markets,'' in {\em International
  Conference on Financial Cryptography and Data Security}, pp.~299--311,
  Springer, 2022.

\bibitem{wu2023towards}
J.~Wu, D.~Lin, Q.~Fu, S.~Yang, T.~Chen, Z.~Zheng, and B.~Song, ``Towards
  understanding asset flows in crypto money laundering through the lenses of
  ethereum heists,'' {\em IEEE Transactions on Information Forensics and
  Security}, 2023.

\bibitem{mosna2023nfts}
A.~Mosna and G.~Soana, ``Nfts and the virtual yet concrete art of money
  laundering,'' {\em Computer Law \& Security Review}, vol.~51, p.~105874,
  2023.

\bibitem{van2023markets}
T.~van~der Linden and T.~Shirazi, ``Markets in crypto-assets regulation: Does
  it provide legal certainty and increase adoption of crypto-assets?,'' {\em
  Financial Innovation}, vol.~9, no.~1, p.~22, 2023.

\end{thebibliography}

\appendix

\section{Cryptoasset Taxonomy in the UK} \label{appendix_uk_taxonomy}

The Consultation paper ``Future Financial Services Regulatory Regime for Cryptoassets'' defines four types of cryptoassets. 

\emph{Exchange Tokens} are a type of cryptoasset that utilizes technology such as \DLT for recording or storing data, and are neither issued nor guaranteed by a central bank or any central authority. \ETH and \BTC are examples of exchange tokens. Sub-types include \emph{stablecoins}, \emph{asset-reference tokens} and \emph{algorithmic tokens}. Stablecoins aim to achieve stability by being pegged to a more stable asset or a basket of assets, such as fiat currency.  Asset-reference tokens reference their value in relation to commodities (i.e., commodity-linked tokens) or other cryptocurrencies (crypto-backed tokens). Algorithmic tokens aim to achieve price stability primarily through an algorithm that adjusts their supply in response to changes in demand and the value of supporting cryptoassets.

\emph{Utility tokens} are cryptoassets that grant digital access to a particular service or application. Unlike security tokens, they do not confer rights or entitlements typically associated with securities, such as ownership or equity, and are not intended to be used as a payment method. Sub-types include governance tokens (protocol voting) and fan tokens (membership voting).

\emph{Security tokens} are cryptoassets that already fulfill the criteria for a specified investment as defined under the RAO, and are therefore subjected to regulation.

\emph{\NFTs} are cryptoassets that represent ownership or proof of authenticity of specific items or pieces of content using \DLT.

Some of these tokens may meet the definition of \emph{e-money tokens}, which are tokens that represent a monetary value stored in electronic form, allowing users to make payments with them. If so, they may fall within the scope of \EMRs. In addition, the UK government does not treat stablecoins as a separate category of cryptoasset but includes them in its existing framework.

\section{Supplementary Tables}

\begin{table}[hbp]
    \resizebox{1\textwidth}{!}{  
    \centering
    \begin{tabular}{c|cccccc}
    \toprule
    Document& Year& \makecell[c]{Authorities}& Regulated Activities& Link \\
    \midrule
     Transfers of Funds Regulation&	2023&	\makecell[c]{The EP and \\the Council}& \makecell[c]{$\diamond$~CASPs are required to \\collect and disclose data \\on cryptoassets transfers} & \href{https://eur-lex.europa.eu/legal-content/EN/TXT/?uri=CELEX\%3A32023R1113}{TFR}\\

     Anti-Money Laundering Directive V&	2018	&\makecell[c]{The EP and \\the Council} & \makecell[c]{$\diamond$~Due diligence, disclosure, \\ data reporting, etc.} &	\href{https://eur-lex.europa.eu/legal-content/EN/TXT/?uri=celex\%3A32018L0843}{AMLD5}\\

     Payment Services Directive	II & 2015	&\makecell[c]{The EP and \\the Council} & $\diamond$~Payment services	&\href{https://eur-lex.europa.eu/legal-content/EN/TXT/?uri=celex\%3A32015L2366}{PSD2} \\

     \makecell[c]{Markets in Financial\\ Instruments Directive II} &	2014	&\makecell[c]{The EP and \\the Council} &	\makecell[c]{$\diamond$~Off-exchange and OTC trading, \\disclosure and reporting, etc.}	& \href{https://eur-lex.europa.eu/legal-content/EN/TXT/?uri=celex\%3A32014L0065}{MiFID2}\\

     Electronic Money Directive II& 2009 &\makecell[c]{The EP and \\the Council} & \makecell[c]{$\diamond$~E-money issuance, \\distribution, redemption, etc.} 	&\href{https://eur-lex.europa.eu/legal-content/en/TXT/?uri=CELEX\%3A32009L0110}{EMD2}\\
     \hline

     Markets in Crypto-Assets Regulation &	2023 &\makecell[c]{The EP and \\the Council}	& \makecell[c]{$\diamond$~Public offerings, the admission\\ to trading, provision of services,\\ market abuse prevention, etc.} &	\href{https://eur-lex.europa.eu/eli/reg/2023/1114/oj}{MiCA}\\

     Digital Finance Package&	2020&	DG FISMA	& \makecell[c]{$\diamond$~Legislative proposals\\ on cryptoassets	}&\href{https://finance.ec.europa.eu/publications/digital-finance-package_en}{DFP} \\

     FinTech Action Plan	& 2018&	DG FISMA	& $\diamond$~Concerns on crypto risks	& \href{https://eur-lex.europa.eu/legal-content/EN/TXT/HTML/?uri=CELEX:52018DC0109\&from=EN}{FAP} \\
     
    \bottomrule
    \end{tabular}
    }
    \caption{Crypoasset Regulation in the EU.}
    \label{tab:EU_reg_overview}
\end{table}

\begin{table}[t]
    \resizebox{1\textwidth}{!}{  
    \centering
    \begin{tabular}{c|cccccc}
    \toprule
    Document& Year& \makecell[c]{Authorities}& Regulated Activities& Link \\
    \midrule
    \makecell[c]{Financial Services and Markets Act} & 2023  &\makecell[c]{FCA,\\PSR}& \makecell[c]{$\diamond$ Investment, trading, issuance, \\payment, financial promotion, etc.  }	& \href{https://www.legislation.gov.uk/ukpga/2023/29/contents}{FSMA}\\
    The AML/CTF Regulations&	2017	&FCA& \makecell[c]{$\diamond$ Customer due diligence, \\disclosure,  reporting, etc.} & \href{https://www.legislation.gov.uk/uksi/2017/692/contents/made}{AML/CTF}\\

    Payment Services Regulations &	2017		&FCA	& $\diamond$~Payment	&\href{https://www.legislation.gov.uk/uksi/2017/752/contents/made}{PSR}\\

    \makecell[c]{Electronic Money Regulations}	&2011	&FCA	& \makecell[c]{$\diamond$~Issuance, management\\ of E-Money}&\href{https://www.legislation.gov.uk/uksi/2011/99/contents/made}{EMR}\\

      \hline
    \makecell[c]{Regulating Cryptoassets \\Phase 1: Stablecoins}& 2023  &FCA	&\makecell[c]{$\diamond$~Stablecoin issuance, \\payments, custody, etc.}	& \href{https://www.fca.org.uk/publications/discussion-papers/dp23-4-regulating-cryptoassets-phase-1-stablecoins}{DP23/4}\\

    \makecell[c]{Update on Plans for the Regulation\\ of Fiat-backed Stablecoins} &	2023	&\makecell[c]{HMT}	&\makecell[c]{$\diamond$~Stablecoin issuance, \\payments, custody, etc.}	& \href{https://assets.publishing.service.gov.uk/media/653a82b7e6c968000daa9bdd/Update_on_Plans_for_Regulation_of_Fiat-backed_Stablecoins_13.10.23_FINAL.pdf}{\makecell[c]{HMT\\Stablecoin}}\\

     \makecell[c]{Future financial services regulatory \\regime for cryptoassets} &	2023	& 	\makecell[c]{HMT}	&\makecell[c]{$\diamond$~crypto issuance, payment, exchange, \\ investment, lending and borrowing,\\ leverage, custody activities.}& 	\href{https://assets.publishing.service.gov.uk/media/653bd1a180884d0013f71cca/Future_financial_services_regulatory_regime_for_cryptoassets_RESPONSE.pdf}{\makecell[c]{HMT\\Cryptoassets}}\\

     \makecell[c]{PS19/22: Guidance on Cryptoassets}	&2019& FCA &	 \makecell[c]{$\diamond$~Crypto issuance, payment,\\ exchange, investment management, \\ financial advising, etc.} & 	\href{https://www.fca.org.uk/publication/policy/ps19-22.pdf}{PS19/22}\\

     Cryptoassets Taskforce &	2018	& \makecell[c]{HMT, \\FCA, \\BoE} &	\makecell[c]{$\diamond$~Laying out the path to \\establish regulatory approach \\to cryptoassets and DLT} & \href{https://assets.publishing.service.gov.uk/media/5bd6d6f0e5274a6e11247059/cryptoassets_taskforce_final_report_final_web.pdf}{Taskforce}\\
    
    \bottomrule
    \end{tabular}
    }
    \caption{Cryptoasset Regulation in the UK.}
    \label{tab:uk_reg_overview}
\end{table}

\end{document}